\pdfoutput=1

\documentclass[12pt]{mn2e}
\usepackage{graphicx}
\usepackage{epsf}
\usepackage{lscape}
\usepackage{longtable}
\usepackage{rotating}
\usepackage{color}

\newcommand{\aap}{A\&A}

\newcommand{\apj}{ApJ}
\newcommand{\apjl}{ApJ}

\newcommand{\mnras}{MNRAS}
\newcommand{\aj}{AJ}

\newcommand{\nat}{Nat}
\newcommand{\mblack}{\em MassiveBlack}
\definecolor{grey}{rgb}{0.7,0.7,0.7}

\begin{document}
\topmargin -0.5in %this is only for astro-ph, uncomment when submitting paper

\title[Confronting predictions of the GSMF]{Confronting predictions of the galaxy stellar mass function with observations at high-redshift}

\author[Stephen M. Wilkins et al.]  
{
Stephen M. Wilkins$^{1}$\thanks{E-mail: stephen.wilkins@physics.ox.ac.uk}, Tiziana Di Matteo$^{1,2}$, Rupert Croft$^{1,2}$, Nishikanta Khandai$^{2,3}$, \newauthor Yu Feng$^{2}$, Andrew Bunker$^{1}$, William Coulton$^{1}$\\
$^1$\,University of Oxford, Department of Physics, Denys Wilkinson Building, Keble Road, OX1 3RH, U.K. \\
$^2$\,McWilliams Center for Cosmology, Carnegie Mellon University, 5000 Forbes Avenue, Pittsburgh, PA 15213, U.S.A.\\
$^3$\,Brookhaven National Laboratory, Department of Physics, Upton, NY 11973, U.S.A.
}
\maketitle 

\begin{abstract}
We investigate the evolution of the galaxy stellar mass function at 
high-redshift ($z\ge 5$) using a pair of large cosmological hydrodynamical 
simulations: {\em MassiveBlack} and {\em MassiveBlack-II}. By combining 
these simulations we can study the properties of galaxies with 
stellar masses greater than $10^{8}\,{\rm M_{\odot}}\,h^{-1}$ and 
(co-moving) number densities of 
$\log_{10}(\phi\, [{\rm Mpc^{-3}\,dex^{-1}}\,h^{3}])>-8$.
Observational determinations of the
galaxy stellar mass function at very-high redshift typically assume a relation
 between the observed UV luminosity
and stellar mass-to-light ratio which is applied to high-redshift samples in
order to estimate stellar masses. This relation can also be measured
from the simulations. We do this, finding two significant differences with
the usual observational assumption: it evolves strongly with redshift
and has a different shape. Using this relation to make a consistent
comparison between galaxy stellar mass functions
we find that at $z=6$ and above the simulation
predictions are in good agreement with observed data over the whole 
mass range. 
Without using the correct UV luminosity
and stellar mass-to-light ratio, the discrepancy would be up to two orders
of magnitude for large galaxies $>10^{10}\,{\rm M_{\odot}}\,h^{-1}$.
At $z=5$, however
the stellar mass function for low mass $<10^{9}\,{\rm M_{\odot}}\,h^{-1}$
galaxies is overpredicted by factors of a few, consistent with 
the behaviour of the UV luminosity function,
and perhaps a sign that feedback in the simulation is not efficient enough 
for these galaxies.
\end{abstract} 


\section{Introduction}

The observational exploration of the high-redshift ($z>2$) Universe has been
driven, over the past 10-15 years, predominantly by deep {\em Hubble Space
  Telescope} (HST) surveys. Deep Advanced Camera for Surveys (ACS)
observations {\em alone} (of the HUDF for example) permitted the
identification of large numbers of galaxies at $z=2-6$ (e.g. 
Bunker et al. 2004, Beckwith et al. 2006, Bouwens et al. 2007). While some galaxies at $z>7$ were
identified using ACS and near-IR Camera and Multi-Object Spectrometer (NICMOS)
observations (e.g. Bouwens et al. 2008) or ground based imaging (e.g. Bouwens
et al. 2008, Ouchi et al. 2009, Hickey et al. 2010) the very-high redshift Universe was only truly
opened up by the installation of Wide Field Camera 3 (WFC3) in 2009. WFC3
near-IR ($1.0-1.6\mu{\rm m}$) observations allow the identification of star
forming galaxies to $z=7-8$ (e.g. Bouwens et al. 2010, Oesch et al. 2010,
Bunker et al. 2010, Wilkins et al. 2010, Wilkins et al. 2011a, Lorenzoni et
al. 2011, Bouwens et al. 2011b) and potentially even to $z\sim 10$ (Bouwens et
al. 2011a, Oesch et al. 2012).

By combining ACS optical and NICMOS or WFC3 near-IR imaging with {\em Spitzer}
IRAC observations it becomes possible to probe the rest-frame UV-optical
spectral energy distributions of galaxies at $z=4-8$ (e.g. Eyles et al. 2005,
Gonzalez et al. 2012). Rest-frame optical photometry is crucial to accurately
determine stellar masses (e.g. Eyles et al. 2007, Stark et al. 2009, Labb{\'e}
et al. 2010, Gonzalez et al. 2011). With a sufficiently large, well defined
sample of galaxies it is possible to study the galaxy stellar mass
demographics, and in particular the galaxy stellar mass function
(e.g. Gonzalez et al. 2011). The galaxy stellar mass function (GSMF) is a
fundamental description of the galaxy population and is defined as the
number density of galaxies per logarithmic stellar mass bin. The first moment of the GSMF corresponds to the cosmic stellar mass density.

Here we use state-of-the-art cosmological hydrodynamical simulations of
structure formation ({\em MassiveBlack} and {\em MassiveBlack-II}) to investigate their predictions of the GSMF and compare it with current
constraints. These runs are large, high resolution
simulations, with more than 65.5 billion resolution elements used in a
box of roughly cubic gigaparsec scales (for {\em MassiveBlack}),
 making it by far the
largest cosmological Smooth Particle Hydrodynamics (SPH) simulation to date 
with “full physics” of galaxy formation
(meaning here an inclusion of radiative cooling, star formation, black hole
growth and associated feedback physics) ever carried out.  The combination of
the two simulations allows us to probe galaxies with stellar masses greater
than $10^{8}\,{\rm M_{\odot}}\,h^{-1}$ and (co-moving) number densities of
$\log_{10}(\phi [{\rm Mpc^{-3}\,dex^{-1}}\,h^{3}])>-8$, a range well matched
with current observations at high-redshift. 

This article is organised as follows: in Section \ref{sec:MB} we
introduce the {\em MassiveBlack} and {\em MassiveBlack-II} simulations. In
Section \ref{sed:GSMF} we explore the predicted evolution of the galaxy
stellar mass function, 
how both the {\em intrinsic} and {\em observed} luminosities
 correlate with the stellar mass-to-light ratio
 and in \S\ref{sed:GSMF.obs} compare galaxy
stellar mass functions to recent
observations. Finally, in Section \ref{sec:c} we present our
conclusions.

Throughout this work magnitudes are calculated using the $AB$ system (Oke \& Gunn 1983). We assume Salpeter (1955) stellar initial mass function (IMF), i.e.: $\xi(m)={\rm d}N/{\rm d}m\propto m^{-2.35}$. 

\section{MassiveBlack and MassiveBlack-II}\label{sec:MB}

\subsection{Simulation runs: {\em Massive Black} and {\em Massive Black-II}}

\begin{table}
\begin{center}
\footnotesize
\caption{Main characterics of  {\em Massive Black} and {\em Massive Black-II} simulations. Both simulations included dark matter, SPH, a multiphase model
for star formation, and a model for black hole accretion and feedback.
   The number of particles
  N$_{\rm part}$ is given, the size of the simulation box  L$_{\rm box}$, the
  gravitational softening length $\epsilon$, the number of cores used 
  $N_{\rm cores}$ and the final redshift $z_f$. 
Both runs were
  started at $z=159$ and used  6 threads/MPI task. For \mblack-II 
 the number of cores and threads used 
was optimized as it progressed.
}
\begin{tabular}{ccccccc}
\hline\hline
  Run & N$_{\rm part}$   &   L$_{\rm box}$ &$\epsilon$& 
  $z_f$  \\
& & (Mpc/h) & (kpc/h)& & \\
  \hline\\
  \mblack & $2\times 3200^3$ & 533 & 5.0&  4.75 & \\
  \mblack-II & $2\times 1792^3$ & 100 & 1.85& 0 \\
\hline\\
\end{tabular}
\normalsize
\label{tab_runs}
\end{center}
\end{table}

Our new simulations (see Table \ref{tab_runs} for the parameters of the simulation) have been performed with the cosmological TreePM-Smooth
Particle Hydrodynamics code {\small P-GADGET}, a {\it hybrid} version of
the parallel code {\small GADGET2} (Springel 2005) which has been
extensively modified and upgraded to run on the new generation of Petaflop
scale supercomputers (e.g. machines like the upcoming BlueWaters at NCSA). The
major improvement over previous versions of {\small GADGET} is in the use of
threads in both the gravity and SPH part of the code which allows the
effective use of multi core processors combined with an optimum number of MPI
task per node.  The {\em MassiveBlack} simulation contains $N_{\it part} = 2
\times 3200^3 = 65.5$ billion particles in a volume of $533\, {\rm Mpc}/h$ on a side
with a gravitational smoothing length $\epsilon = 5.0\, {\rm kpc}/h$ in comoving
units. The gas and dark matter particle masses are $m_{\rm g} = 5.7 \times
10^7 M_{\odot}$ and $m_{\rm DM} = 2.8 \times 10^8 M_{\odot}$ respectively.  The
simulation has currently been run from $z=159$ to $z=4.75$ (beyond our
original target redshift of $z=6$). For this massive calculation it is
currently prohibitive to push it to $z=0$ as this would require an
unreasonable amount of computational time on the world's current fastest
supercomputers. The simulated redshift range probes early structure formation
and the emergence of the first galaxies and quasars. 

 {\em MassiveBlack-II} (see Khandai et al. {\em in-prep} for an overview) is
a smaller volume but the mass and spatial resolution are 
better than {\em MassiveBlack} by a factor of 25
and 2.7 respectively. The smaller volume means that a smaller part of the high
mass function can be sampled and that in the mass range where it overlaps with
{\em MassiveBlack} it can be used to check for convergence
as well as to extend
our predictions towards the low mass end. This is the largest volume ever run at
this resolution with a final redshift of 
$z=0$.

These runs contain gravity and hydrodynamics but also extra physics (subgrid modeling)
for star formation (Springel \& Hernquist 2003), black holes and associated feedback
processes (Di Matteo et al.  2008, Di Matteo et al. 2012).  The cosmological parameters
used were: the amplitude of mass fluctuations, $\sigma_8=0.8$, spectral index,
$n_s = 0.96$, cosmological constant parameter $\Omega_{\Lambda}= 0.74$, mass
density parameter $\Omega_m = 0.26$ , baryon density parameter $\Omega_b =
0.044$ and $h=0.72$ (Hubble's constant in units of $100 \mathrm{km\:s}^{-1}
\mathrm{Mpc}^{-1}$; WMAP5) for {\em MassiveBlack}. For {\em MassiveBlack-II} we instead used
 $\Omega_{\Lambda}= 0.725$,
and $\Omega_m = 0.275$ (according to WMPA7).

Catalogues of galaxies are made from the simulation outputs by first using 
a friends-of-friends groupfinder and then applying the {\sc SUBFIND} algorithm
(Springel 2001) to find gravitationally bound subhalos. The stellar
component of each subhalo consists of a number of star particles, each labelled
with a mass and the redshift at which the star particle was created. 

To generate the spectral energy distribution (SED), and thus broad-band
photometry, of each galaxy we sum the SEDs of each star particle (weighted by
the particle mass). The SED of each star particle is generated using the {\sc
  Pegase.2} stellar population synthesis (SPS) code (Fioc \& Rocca-Volmerange 1997,1999) taking account of
their ages and metallicities. Nebula (continuum and line) emission is also added to each star
particle SED, though this has a
negligible effect on the UV photometry considered in this work. In addition we
apply a correction for absorption in the intergalactic medium (IGM) using the
standard Madau et al. (1995) prescription (though again this has a negligible effect
on this work). Throughout this work we measure the broad-band UV luminosity
using an idealised rest-frame top-hat filter at $\lambda=1500\pm 200{\rm
  \AA}$. A rest-frame filter is chosen to allow a consistent comparison
between samples at different redshifts. The shape of this filter is selected
for convenience, but closely reflects the profile of near-IR bandpasses which
are available to measure the rest-frame UV flux at high-redshift.

We note that our work is complementary to the recent simulation predictions
of the galaxy stellar mass functions of Jaacks et al. (2012), who compare
results for a suite of smaller simulations to the Gonzalez et al. (2011, hereafter G11)
observational data. Our work differs in extending to a lower redshift,
correcting for the effect of an evolving ratio of UV luminosity to mass to light
ratio, and also for the inclusion of supermassive black hole formation and
feedback in our simulations. We discuss the Jaacks et al. (2012) results
further below.

\section{The Galaxy Stellar Mass Function}\label{sed:GSMF}

Measuring the GSMF from outputs of the {\em MassiveBlack} and 
{\em MassiveBlack-II} simulations is straightforward, given that the
total masses of star particles in each galaxy are known. Before making a
comparison to observational data, however, we must 
remember that observed  UV luminosities were used (e.g. by Gonzalez et al. 2011, hereafter G11) to 
compute the published observed GSMFs. 
This means examining the relationship
between UV luminosity and stellar mass to light ratio in the simulation
and using this information in our comparison to observations.
In this section, we do this, after first presenting the GSMF measured
directly from the simulations. 

\subsection{Galaxy Stellar Mass Function from simulations}

\begin{figure}
\centering
\includegraphics[width=22pc]{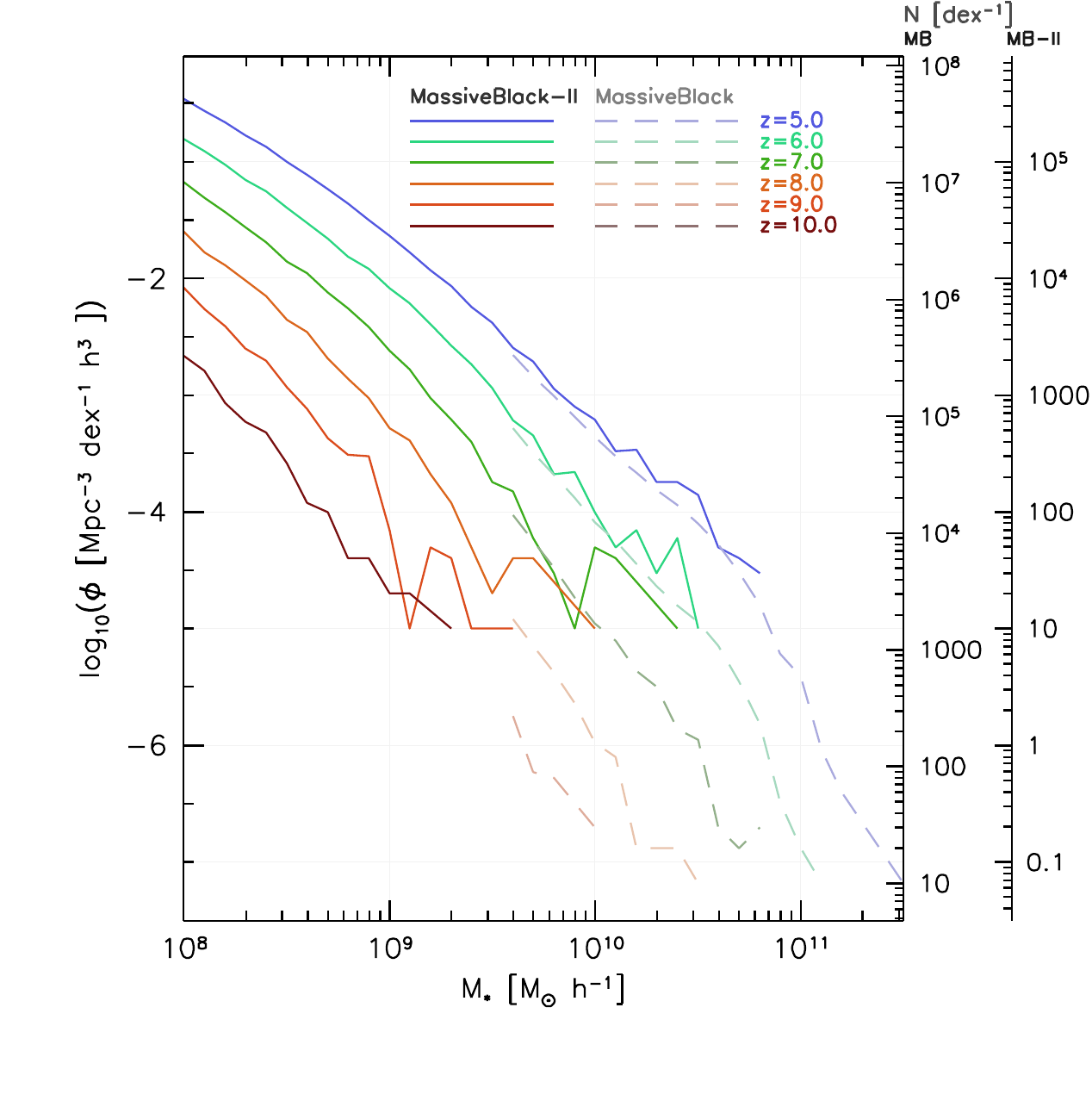}
\caption{The galaxy stellar mass function measured from the {\em MassiveBlack}
  (dashed lines) and {\em MassiveBlack-II} (solid lines) simulations for
  $z\in\{5,6,7,8,9,10\}$. The two right hand axes show the number of galaxies
  in the {\em MassiveBlack} and {\em MassiveBlack-II} volumes.}
\label{fig:GSMF}
\end{figure}

The evolution of the $>10^{8}\,{\rm M_{\odot}}\,h^{-1}$ galaxy stellar mass
function from $z=10\to 5$ predicted by {\em MassiveBlack} and {\em
  MassiveBlack-II} is shown in Fig. \ref{fig:GSMF}. The shape of the simulated
GSMF is a declining distribution with mass and, at least at $z=5$, exhibits a
sharp cut off at high-masses. Values of the number density $\phi$ are also tabulated in Table \ref{tab:gsmf} in various logarithmic mass intervals.

Figure \ref{fig:GSMF} also demonstrates the evolution in the normalisation of
the GSMF. At $z=10$ there are only $\sim 500$ galaxies with stellar masses
$>10^{8}\,{\rm M_{\odot}}\,h^{-1}$ in the {\em MassiveBlack-II} volume
($10^{6}\,{\rm Mpc^{3}\,}h^{-3}$), while at $z=5$ this has increased to $\sim
135,000$ ($\times 270$). The shape of the GSMF also evolves strongly; while
the number of galaxies with masses $>10^{8}\,{\rm M_{\odot}}\,h^{-1}$
increases by a factor of $\times 270$ from $z=10\to 5$ the number of galaxies
with masses $>10^{10}\,{\rm M_{\odot}}\,h^{-1}$ increases by a factor of
$\times 5000$.

The evolution of the simulated galaxy stellar mass function is stronger than
that exhibited by the UV luminosity function. This
reflects the fact the average UV mass-to-light ratio of galaxies also
increases $z=10\to 5$ (as demonstrated in Section \ref{sec:UVMTOL}).

\begin{table*}
\begin{center}
\footnotesize
\caption{The number density (in units of ${\rm Mpc^{-3}\,dex^{-1}\,}h^{3}$) of galaxies in various logarithmic mass intervals ($[9.5,10.0)\equiv 9.5\le\log_{10}(M)<10.0$, where $M$ has units ${\rm M_{\odot}\,}h^{-1}$) for $z\in\{5,6,7,8,9,10\}$. Where there are no objects within the mass interval the number density is replaced by an upper limit corresponding to $n<1$ (i.e. $\phi<1/V$).}
\begin{tabular}{lcccccc}
Mass Interval  & \multicolumn{6}{c}{$\log_{10}(\phi\, {\rm [Mpc^{-3}\,dex^{-1}\,}h^{3}{\rm]})$} \\
 \hline
$\log_{10}({\rm [M_{\odot}\,}h^{-1}{\rm ]})$ & $z=5$ & $z=6$ & $z=7$ & $z=8$ & $z=9$ & $z=10$ \\
 \hline\hline
 &\multicolumn{6}{c}{{\em MassiveBlack}  $\,\,\,{\rm Volume}=(533\,{\rm Mpc}\,h^{-1})^{3}$} \\
 \hline
$[9.5,10.0)$ & $-2.85$ & $-3.50$ & $-4.25$ & $-5.13$ & $-6.07$ & $-7.88$ \\ 
$[10.0,10.5)$ & $-3.69$ & $-4.46$ & $-5.35$ & $-6.43$ & $-7.88$ & $<-8.18$ \\ 
$[10.5,11.0)$ & $-4.55$ & $-5.46$ & $-6.80$ & $-7.88$ & $<-8.18$ & $<-8.18$ \\ 
$[11.0,11.5)$ & $-6.23$ & $-7.88$ & $<-8.18$ & $<-8.18$ & $<-8.18$ & $<-8.18$ \\ 
$[11.5,12.0)$ & $<-8.18$ & $<-8.18$ & $<-8.18$ & $<-8.18$ & $<-8.18$ & $<-8.18$ \\ 

\hline
& \multicolumn{6}{c}{{\em MassiveBlack-II}  $\,\,\,{\rm Volume}=(100\,{\rm Mpc}\,h^{-1})^{3}$} \\
 \hline
$[8.0,8.5)$ & $-0.70$ & $-1.06$ & $-1.46$ & $-1.92$ & $-2.44$ & $-3.03$ \\ 
$[8.5,9.0)$ & $-1.27$ & $-1.69$ & $-2.15$ & $-2.69$ & $-3.31$ & $-4.06$ \\ 
$[9.0,9.5)$ & $-1.95$ & $-2.42$ & $-3.01$ & $-3.71$ & $-4.47$ & $-5.10$ \\ 
$[9.5,10.0)$ & $-2.76$ & $-3.37$ & $-4.12$ & $-4.80$ & $-5.70$ & $<-6.00$ \\ 
$[10.0,10.5)$ & $-3.57$ & $-4.27$ & $-4.74$ & $-5.70$ & $<-6.00$ & $<-6.00$ \\ 
$[10.5,11.0)$ & $-4.40$ & $<-6.00$ & $<-6.00$ & $<-6.00$ & $<-6.00$ & $<-6.00$ \\ 
\hline
\end{tabular}
\normalsize
\label{tab:gsmf}
\end{center}
\end{table*}

\subsection{Observational Estimation of the Galaxy Stellar Mass Function}

By combining HST optical and near-IR observations (from ACS and NICMOS or
WFC3) with {\em Spitzer} IRAC photometry it is possible to measure the
rest-frame UV-optical spectral energy distributions of high-redshift
galaxies. Rest-frame optical photometry is vital to determine accurate stellar
masses. Several studies have recently attempted to measure the stellar masses
of high-redshift Lyman-break selected galaxies (e.g. Eyles et al. 2007, Stark
et al. 2009, Labb{\'e} et al. 2010, Gonzalez et al. 2011). With a sufficiently
large sample and a handle on the incompleteness issues it is also possible to
study the galaxy stellar mass function (e.g. Stark et al. 2009, Labb{\'e} et
al. 2010, Gonzalez et al. 2011).

To understand how to make simulation predictions
it is useful to examine exactly
how the G11 GSMF is constructed. The G11 study draws a sample of galaxies from
the {\em observed} UV luminosity functions (LFs) at $z\in\{3.8,5.0,5.9,6.8\}$
(using Bouwens et al. 2007, 2011). These UV luminosities are converted into
stellar masses using the {\em observed} UV Luminosity ($L_{1500,obs}$) -
stellar mass-to-light ratio ($M/L_{1500,obs}$) distribution measured at $z\sim
4$. This relation is fairly well fit by a power law\footnote{Though the power
  law fit is not used to determine the GSMF.}, such that
$M/L_{1500,obs}\propto L^{0.7}$ (i.e. the stellar mass-to-light ratio
increases with observed UV luminosity). While this relation is calibrated at
$z=4$ G11 note that that it appears to fit observations of stellar masses and
luminosities at $z\in\{5,5.9\}$. However at these redshifts the sample sizes are
small (78 and 28 galaxies at $z\sim 5$ and $z\sim 6$ respectively) and there is a
large degree of scatter.

\subsection{The relation between UV luminosity and the stellar mass-to-light ratio in simulations}\label{sec:UVMTOL}

As noted above, the G11 study uses the distribution of stellar masses and UV
luminosities measured at $z\sim 4$ to effectively convert the observed UV
luminosity function into a galaxy stellar mass function.
To make a proper simulation prediction we must take into account
any difference between the
relation between UV luminosity and the stellar mass-to-light ratio
used by G11 and that in the simulations.

\begin{figure}
\centering
\includegraphics[width=20pc]{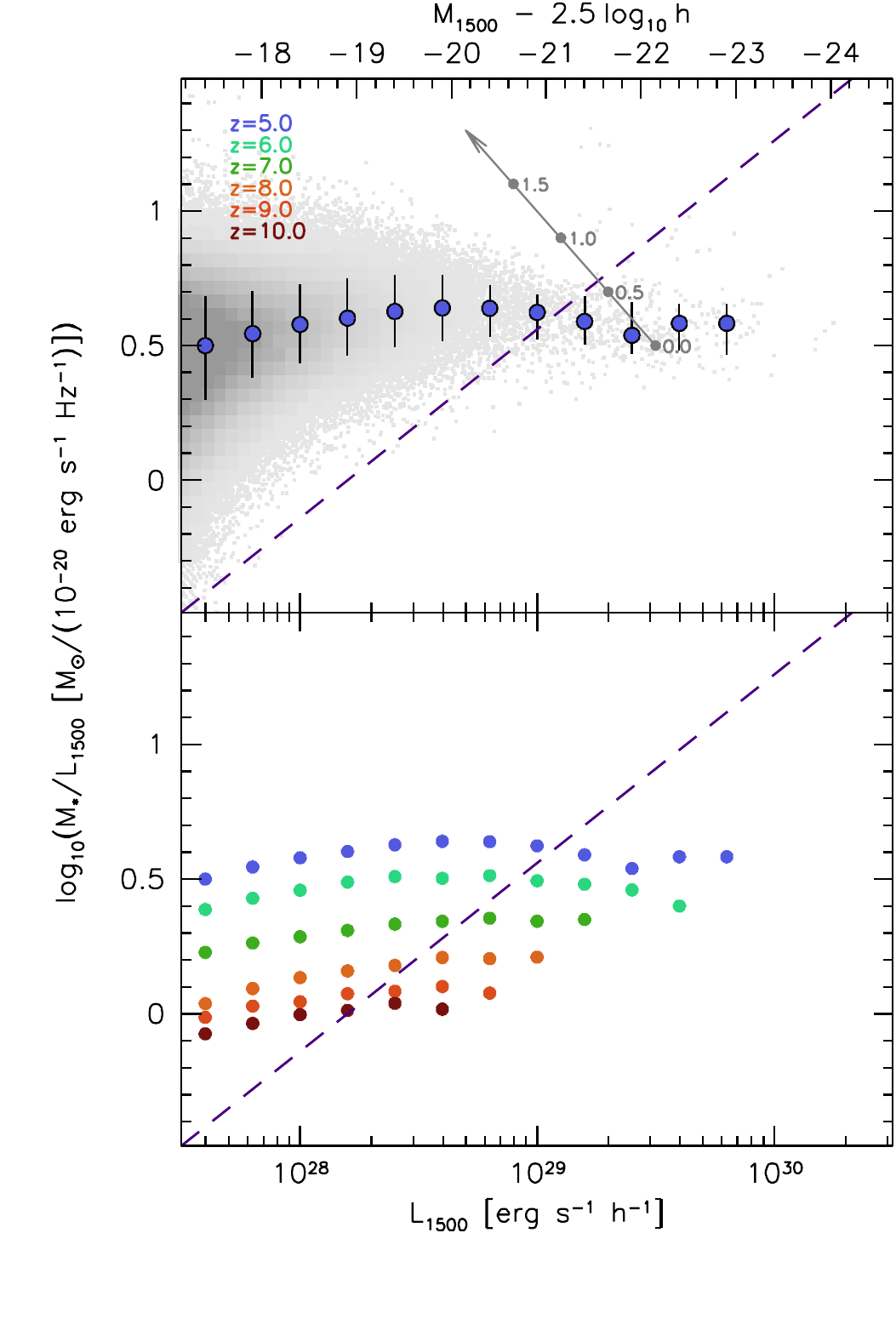}
\caption{The relationship between the {\em intrinsic} UV luminosity and
  stellar mass-to-light ratio at $z\in\{5,6,7,8,9,10\}$ predicted from {\em
    MassiveBlack-II}. In both panels the points denote the median value of the
  mass-to-light ratio in each luminosity bin. In the upper-panel the 2-d
  histogram shows the density of sources on a linear scale and the error bars
  show the range encompassing the central 68.2$\%$ of galaxies. The arrow in
  the upper-panel shows the effect of dust attenuation (the labels denote
  values of $A_{1500}$). The dashed line in both panels shows
  $M/L_{1500}\propto L^{0.7}$ which provides a good fit to the distribution
  used by Gonzalez et al. (2011) to determine stellar masses from {\em
    observed} UV luminosities.}
\label{fig:L_UVMTOL}
\end{figure}

Figure \ref{fig:L_UVMTOL} shows the relationship between the intrinsic UV
luminosity ($L_{1500}$) and mass-to-light ratio ($M/L_{1500}$) at
$z\in\{5,6,7,8,9,10\}$ predicted by {\em MassiveBlack-II}. This relationship
is (over the full mass range) approximately flat (i.e. the {\em intrinsic}
stellar mass-to-light ratio is constant) and is significantly different from
the $M/L_{1500,obs}\propto L^{0.7}$ relation found by G11.
Jaacks et al. (2012) plotted the rest frame UV magnitude against stellar mass
in their simulations, also finding a flatter relationship that than used by
G11. The lower-panel of Fig. \ref{fig:L_UVMTOL} shows that the
relationship between the intrinsic UV luminosity and stellar mass-to-light
ratio also varies strongly with redshift, increasing by $0.6\,{\rm dex}$ from
$z=10\to 5$.

It is also interesting to note
from Figure \ref{fig:L_UVMTOL}
 that it appears the {\em intrinsic} UV luminosity
of galaxies with $L_{1500}>10^{28}\,{\rm erg\,s^{-1}}\,h^{-1}$ can {\em alone}
be used to estimate the stellar mass with an accuracy of $\approx 50\%$. This
contrasts sharply with the low-redshift Universe where star formation has terminated in many systems (particularly massive ellipticals) rendering the UV luminosity to be negligible. The strong correlation between UV luminosity and stellar mass reflects the fact that
virtually all galaxies at high-redshift (in the {\em MassiveBlack} and {\em MassiveBlack-II} simulations) continue to actively
form stars.

\subsection{The effect of dust attenuation}

\begin{figure}
\centering
\includegraphics[width=20pc]{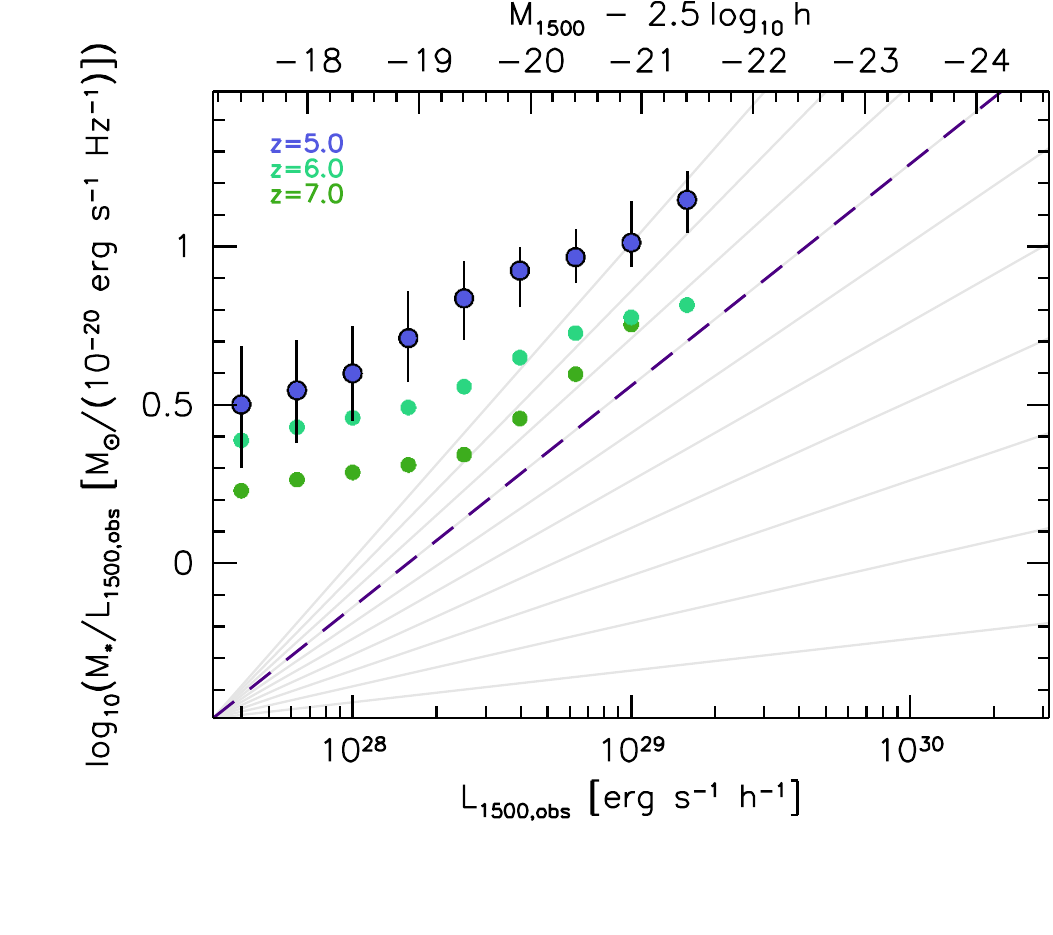}
\caption{The relationship between the dust attenuated (observed) UV luminosity
  and stellar mass-to-light ratio at $z\in\{5,6,7\}$ predicted from {\em
    MassiveBlack-II} and using Bouwens et al. (2012) to relate dust
  attenuation to the observed UV luminosity. The points denote the median
  value of the mass-to-light ratio in each bin while the vertical error bars
  (at $z=5$) denote the 68.2$\%$ confidence interval. The diagonal lines
  denote $M/L_{1500,obs}\propto L^{\gamma}$ for
  $\gamma=\{0.1,0.2,...,1.0\}$. The dashed line denotes $\gamma=0.7$.}
\label{fig:L_UVMTOL_obs}
\end{figure}

The G11 relation is however based on the {\em observed} (i.e. dust attenuated
luminosities) as opposed to the {\em intrinsic} luminosities (as used in
Fig. \ref{fig:L_UVMTOL}). Attenuation due to dust both decreases the UV
luminosity (i.e. $L_{1500,obs}<L_{1500}$) and {\em increases} the stellar
mass-to-light ratio (i.e. $M/L_{1500,obs}>M/L_{1500}$) relative to their
intrinsic values. A positive correlation between luminosity and dust
attenuation would then introduce a positive correlation between
$M/L_{1500,obs}$ and the observed UV luminosity.

The measurement of dust attenuation at high-redshift is challenging. Far-IR
observations, and optical emission lines, are generally inaccesible for the
bulk of the galaxy population at high-redshift leaving only the UV continuum
slope $\beta$ as a diagnostic (e.g. Meurer et al. 1999, Wilkins et
al. 2012a, Wilkins et al. {\em submitted}). A number of recent studies have attempted to constrain the relationship between $\beta$
and the observed UV luminosity at high-redshift though with some conflicting results (e.g. Stanway, McMahon, \& Bunker 2005, Bouwens
et al. 2009, Wilkins et al. 2011b, Dunlop et al. 2012, Bouwens et al. 2012, Finkelstein et al. 2012). Bouwens
et al. (2009), Wilkins et al. (2011b) and Bouwens et al. (2012) an increase in
$\beta$ with observed luminosity. Dunlop et al. (2012) and Finkelstein et
al. (2012) on the other hand found little evidence of variation of $\beta$
with luminosity (see Wilkins et al. {\em submitted} for a detailed comparison).

Adopting the relationship(s)\footnote{If a luminosity invariant dust
  correction was assumed the shape of the observed UV luminosity -
  mass-to-light ratio relation would remain the same (though the average
  observed mass-to-light ratio would increase).} between $\beta$ and
luminosity found by Bouwens et al. (2012) and utilising the Meurer et
al. (1999) calibration (between the observed UV continuum slope $\beta$ and UV attenuation) we can determine the relationships between the observed
UV luminosity ($L_{1500,obs}$) and observed mass-to-light ratio at
$z\in\{5,6,7\}$ as predicted by {\em MassiveBlack} and {\em MassiveBlack-II}. These are shown in Fig. \ref{fig:L_UVMTOL_obs}. The most
significant change (relative to that found for the intrinsic luminosities and mass-to-light ratios) is that the relationship between $L_{1500,obs}$ and
$M/L_{1500,obs}$ is no longer approximately constant but is instead strongly positively correlated, at least at $M_{1500,{\rm obs}}<-19.5$
. At $M_{1500,obs}<-19.5$ the slope of this relation is $\gamma=0.5-0.8$ (where $\gamma$ is defined such that $M/L_{1500,obs}\propto L^{\gamma}$)
(c.f. $\gamma=0.7$ found by G11 at $z=4$). This suggests the physical cause of the
strong observed correlation between UV luminosity and mass-to-light ratio is
caused almost solely by the correlation of dust attenuation with luminosity. At lower-luminosities the relation flattens ($\gamma <0.2$). This arises due to the diminishing effect of dust at lower-luminosities,
i.e. the $L_{1500,obs}$ - $M/L_{1500,obs}$ begins to reflect the (virtually flat)
intrinsic relation.

\begin{figure}
\centering
\includegraphics[width=20pc]{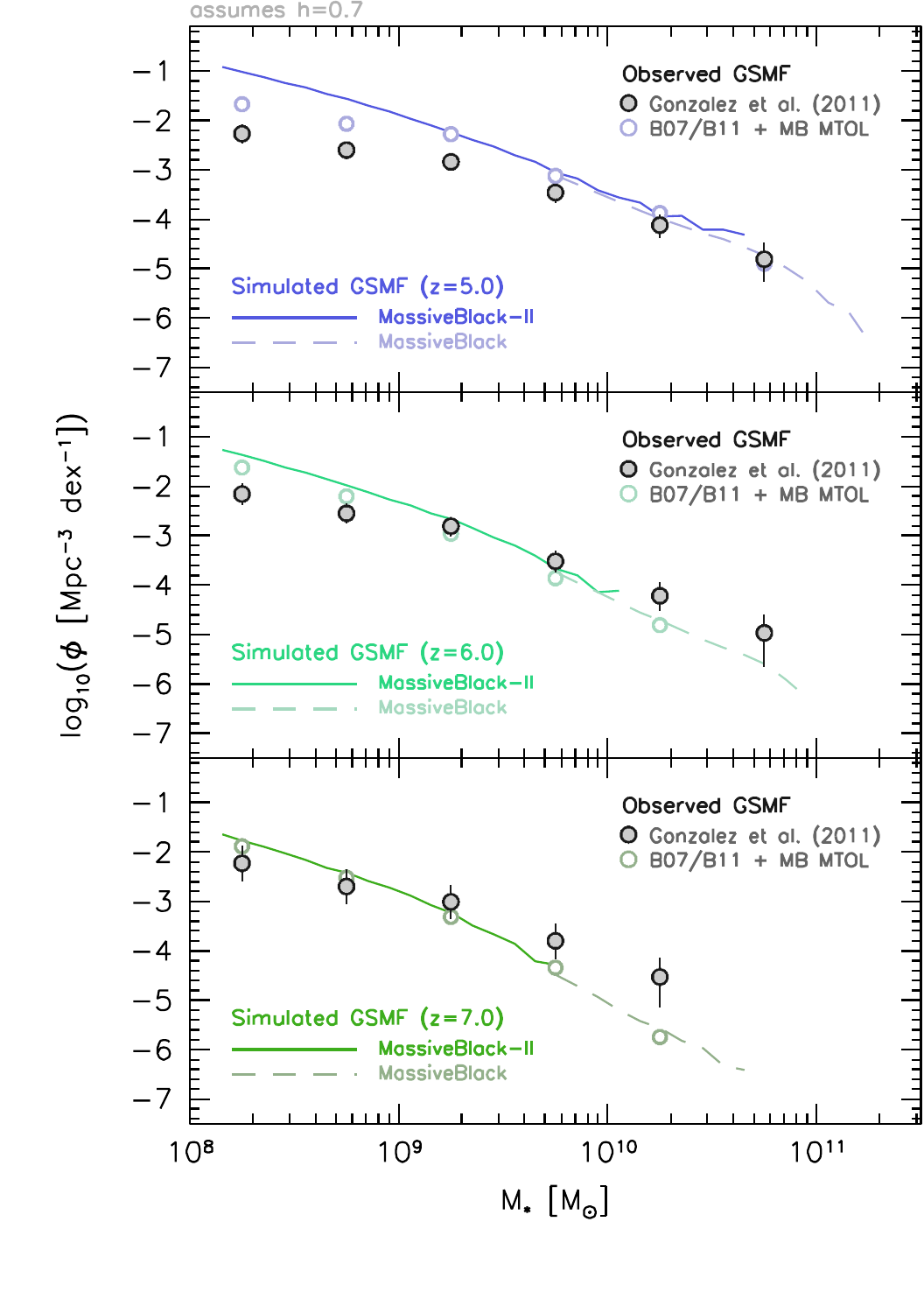}
\caption{The galaxy stellar mass function predicted by  {\em MassiveBlack} (dashed lines) and {\em MassiveBlack-II} (solid lines) compared with observations at $z\in\{5,6,7\}$ (top, middle, and bottom panels respectively). 
 The open symbols in each panel show the prediction for the GSMF 
using the Bouwens et al. (2007, 2011) observed UV LF and a relationship between stellar mass and luminosity derived from {\em MassiveBlack-II}.
The filled grey points show the  GSMF  from Gonzalez et al. (2011),
which was estimated using a non-evolving relationship between UV luminosities
and stellar mass to light ratios. Note that the units now implicitly assume $h=0.7$.}
\label{fig:GSMF_obs}
\end{figure}

\subsection{Comparison with observations}\label{sed:GSMF.obs}

We are now in position to compare
the {\em MassiveBlack}
and {\em MassiveBlack-II} results
to observations. We follow a procedure similar to G11 but
using the simulated relation between UV luminosity and mass-to-light ratio. We
construct a volume limited sample
(referred to below as
 ``B07/B11+MB MTOL'') of galaxy UV luminosities using the Bouwens
et al. (2007, 2011) observed UV luminosity functions. We then convert the {\em
  observed} UV luminosity of each galaxy to a stellar mass using the relation
between luminosity and stellar mass-to-light ratio ($M/L_{1500,obs}$)
predicted by the {\em MassiveBlack-II} simulation (combined with the empirical dust correction described above) and construct a galaxy
stellar mass function. These galaxy stellar mass functions are shown
at $z\in\{5,6,7\}$ in Fig. \ref{fig:GSMF_obs}.
We also show in 
Fig. \ref{fig:GSMF_obs} the GSMFs
predicted by {\em MassiveBlack}/{\em MassiveBlack-II} and those determined by
Gonzalez et al. (2011, {\em hereafter} G11) at $z\in\{5,6,7\}$ (the $z\sim 7$
GSMF comes from Labb{\' e} et al. 2010 but is also presented in G11).

From an examination of 
 Fig. \ref{fig:GSMF_obs} it is clear that the
B07/B11+MB MTOL sample
shows a much closer correspondence to the
simulations compared to G11. This essentially reflects the good overall
agreement between the simulated UV LF and the observations, at least at high-luminosities.
The flattening of the relation between $L_{1500,obs}$ and the mass-to-light
ratio at low-luminosities does go some way to explaining the difference in the
shape of the simulated and observed galaxy stellar mass functions. More
importantly however is the strong redshift evolution: from $z=5\to 7$ the
calibration relating the the observed UV luminosity to the mass-to-light ratio
decreases by $0.3-0.5\,{\rm dex}$ (depending on the luminosity). Because the
G11 study assumed no redshift evolution (instead utilising a calibration based
on observations at $z\sim 4$ to convert UV luminosities to stellar masses at
$z=4-7$) this would cause the stellar masses to be overestimated. Because the
GSMF declines to high-masses this would cause the number density of sources at
any mass to overestimated.

We also note from Fig. \ref{fig:GSMF_obs} that at $z=5$ (and to a lesser extent at $z=6$) this process does
 not fully reconcile the GSMF
at low-masses. At $z=5$ {\em MassiveBlack-II}
over-predicts the faint-end of the UV luminosity function relative to the
observations of Bouwens et al. (2007) by around a factor $\times 5$ at
$M_{1500}=-18$. This is difficult to reconcile observationally without
requiring the application of a much larger completeness correction. It
 therefore suggests that the discrepancy has its roots
in the {\em MB}/{\em MB-II} modelling assumptions. This disagreement occurs in low mass galaxies which are much less affected by
AGN feedback and hence more sensitive to the details of the star formation
model and stellar feedback.  For example our model does not include any
treatment of the molecular gas component such as in e.g. Krumholtz and Gnedin
(2011) which would tend to suppress star formation rates in lower mass
galaxies. However recent simulations of isolated galaxies (e.g. Hopkins,
Quataert \& Murray 2012) have shown that, in the presence of feedback,
restricting star formation to molecular gas or modifying the cooling function has very little effects on the star formation rates. By contrast changing feedback mechanism or associated efficiencies translates in large differences in final stellar mass densities.  Based on these recent results (albeit on idealized simulations) we are prone to interpret our discrepancy at the low mass end to details in the stellar feedback model (and in particular to its efficiency which may be too low).

Comparing to the simulation results of Jaacks et al. (2012) (which do not include AGN modelling), we see that
a similar sign to the disagreement with observations at low mass. At the high mass end,
we have shown that correcting for the UV luminosity-mass to light ratio
assumed brings the observations and simulations into agreement, and this
would also be likely to work for the Jaacks et al. results.
Finally, it is also worth noting that the G11 GSMF evolves only very mildly
from $z=5\to 7$. Indeed, the stellar mass density (which is the first moment
of the GSMF) of galaxies with $>10^{8}\,{\rm M_{\odot}}$ is virtually flat
$z=5\to 7$. This is surprising given that all the galaxies contributing to the
GSMF at these redshifts/masses are likely actively forming stars (by virtue of
being UV selected) and suggests either the high-redshift GSMF is overestimated
or the lower-redshift GSMF underestimated.

\section{Conclusions}\label{sec:c}

We have investigated the high-redshift ($z=5-10$) evolution of the galaxy
stellar mass function (GSMF) using a pair of large cosmological hydrodynamic
simulations {\em MassiveBlack} and {\em MassiveBlack-II}. Over the redshift
range $z=10\to 5$ we find both the normalisation and shape of the GSMF evolves
strongly with the number density of massive galaxies ($>10^{8}\,{\rm
  M_{\odot}}$) increasing by a factor of around $\times 300$.

By combining {\em Hubble Space Telescope} optical and near-IR observations
(from ACS, NICMOS and WFC3) with near-IR IRAC photometry from the {\em Spitzer
  Space Telescope} it is possible to identify and measure the stellar masses
of galaxies at very-high redshift, and thus constrain the GSMF (e.g. Gonzalez
et al. 2011). While the simulated GSMF at $z=5$ provides reasonable agreement
with the Gonzalez et al. (2011) observations at $>10^{9.5}\,{\rm M_{\odot}}$,
at low-masses and at $z>5$ there is a significant discrepancy. The disagrement
at low-masses at $z=5$ is also reflected in the UV luminosity
function (LF) at low-luminosities. However,
at $z>5$ the discrepancy appears to arise due to a difference in the assumed
relationship between the observed UV luminosity and mass-to-light
ratio. Gonzalez et al. (2011) applies a relationship calibrated at $z\sim 4$,
however we find that the relation, while having a similar form (i.e. that the
mass-to-light ratio is positively correlated with the observed UV luminosity),
evolves strongly with redshift. Applying a calibration based on the simulated
distribution of UV luminosities and stellar masses to the observed UV
luminosity functions yields galaxy stellar mass functions which closely
reflect those predicted by the simulations. This simply reflects the
good agreement between the observed and simulated intrinsic UV luminosity
functions.

\subsection*{Acknowledgements}
We would like to thanks Joseph Caruana and the anonymous referee for useful discussions and suggestions. SMW and AB acknowledge support from the Science and Technology Facilities Council. RACC thanks the Leverhulme
Trust for their award of a Visiting Professorship at the University of Oxford. WRC acknowledges support from an Institute of Physics/Nuffield Foundation funded summer internship at the University of Oxford. The simulations were run on the Cray XT5 supercomputer Kraken at the National Institute for Computational Sciences. This research has been funded by the National Science Foundation (NSF) PetaApps program, OCI-0749212 and by NSF AST-1009781.

\bsp

\end{document}